\documentclass[12pt]{article}

\usepackage{amsmath,amssymb}
  \usepackage{authblk}
\usepackage{hyperref}  
\usepackage{cite}

\usepackage{enumitem}  
 
\usepackage[usenames, dvipsnames]{color} 
\usepackage{graphicx}     
\usepackage{subcaption}
\usepackage[utf8]{inputenc}
\usepackage{url}
 \usepackage{xcolor}

\def\beq{\begin{equation}}
\def\eeq{\end{equation}}

\begin{document}

\thispagestyle{empty}

\centerline{{\Large \bf  Black hole entropy and long strings}} 
\vskip1.35truecm
%Black hole entropy and long strings

\begin{center}
{\large Erik P. Verlinde$^{1}$  \quad  and \quad Manus R. Visser$^{2}$}\\
\vskip .75truecm
$^1${\it Institute for Theoretical Physics,  University of Amsterdam, \\ 1090 GL Amsterdam, The Netherlands}\\[1mm]
$^2${\it Department of Theoretical Physics, University of Geneva, 24 quai Ernest-Ansermet, 1211 Gen\`{e}ve 4, Switzerland}\\
\vskip .75truecm
  {\tt e.p.verlinde@uva.nl}$\qquad$ {\tt manus.visser@unige.ch}

\vskip 1cm

\end{center}
\vskip 1.5cm

\begin{abstract}

 We discuss whether black hole entropy counts short   or long range microstates in quantum gravity. In brick wall and induced gravity models the entropy arises due to short distance correlations   across the event horizon cut off at the Planck length. However,    the energy of these short range degrees of freedom is too high compared to the   black hole energy. We argue that   the long string phenomenon, which naturally appears in matrix quantum mechanics,  resolves this issue by lowering the  excitation energy per degree of freedom.   This mechanism also   reduces the total  number of microscopic degrees of freedom in a given volume, leading to a correct estimate of the Bekenstein-Hawking formula for black hole entropy. %Thus, it seems   the long string phenomenon is a necessary ingredient of quantum gravity.% that the microstates are long stringsby reducing the number of UV degrees of freedom and   lowering their excitation energy. We derive the black hole entropy from a 
 
%  In contrast, in black hole-string transition the entropy count is dominated by long string microstates. We present a long string model where   black hole entropy follows from the Cardy formula for  a two-dimensional symmetric product CFT. We speculate the two models are consistent through the Reeh-Schlieder theorem.

%We argue that black hole entropy counts the number of IR degrees of freedom in a microscopic model of gravity. We show the relation with the long string phenomenon in symmetric product CFTs and derive the Bekenstein-Hawking entropy from the associated Cardy formula. 
 
\end{abstract}

\thispagestyle{empty}

\vskip 2cm

\begin{center}
{ \it Essay written for the Gravity Research Foundation 2022 Awards \\for Essays on Gravitation; submitted March 31, 2022.}
\end{center}

\newpage 
\setcounter{page}{1}

Black holes are true thermodynamic objects with an entropy proportional to the horizon area, $S_{BH}=A/4 G \hbar$,  and a temperature given by $T_H =   \kappa \hbar / 2\pi$, where $\kappa$ is the surface gravity \cite{Bekenstein:1972tm,Bekenstein:1973ur,Hawking:1974sw}.\footnote{We set $c=1$  but we  keep $\hbar$ and $G$ explicit in our equations.} Just like ordinary thermodynamic entropy, black hole entropy should have a statistical mechanical origin. It is not known, however, even for the simple  case of asymptotically flat Schwarzschild black holes what the microstates   are that give rise  to their entropy. Since the entropy is independent of the details of the quantum gravity theory, it is expected that the microstates are universal  to leading order in $1/G$. However,  fifty years after Bekenstein's   proposal \cite{Bekenstein:1972tm} that black holes carry entropy there is still  no consensus on  the description and nature of the microstates. In particular, it is not clear whether the microscopic degrees of freedom responsible for the black hole entropy are localized within a Planck length of the horizon, or are non-locally distributed over scales comparable to the horizon size.

%INTRODUCTION Black hole entropy provides a window into the number and nature of microstates in quantum gravity. It suggests that the dimension of the Hilbert space in quantum gravity in a certain spatial region is bounded by the area of its boundary. 

The  proposals for the black hole microstates  can be   grouped into roughly two   categories.  %there are two main explanations of black hole entropy in the literature. 
On the one hand, in brick wall \cite{tHooft:1984kcu} and induced gravity models~\cite{Sakharov:1967pk,Jacobson:1994iw,Frolov:1996aj}   the quantum correlations between the exterior and interior of the black hole horizon   account for the black hole entropy. Also in loop quantum gravity   the entropy arises due to the discreteness of the horizon geometry at the Planck scale  \cite{Rovelli:1996dv}.  %live close to the horizon. 
On the other hand, in string theory the entropy of (near)-extremal black holes   is derived from the Boltzmann entropy by counting the   degeneracy of microstates of a D-brane configuration \cite{Strominger:1996sh}. D-branes are non-perturbative objects which probe long distances at high energies in their world-volume theory. Further, in the  fuzzball program \cite{Mathur:2005zp} and in black hole-string correspondence \cite{Susskind:1993ws,Horowitz:1996nw,Chen:2021dsw} the microstates of black holes are allowed to fluctuate at scales of the   horizon size. %In the first model black hole entropy is due to short range correlated fluctuations whereas in the second theory the entropy follows from counting the number of non-local microstates. 
At first sight,  the two different proposals seem at odds with each other, as the microstates have   completely different length scales, but they might be two different sides of the same coin.  

In fact, in the AdS/CFT duality    the  entropy of black holes is either von Neumann entropy or Boltzmann entropy depending on whether the black hole is one- or two-sided. %the AdS/CFT correspondence both proposals explain black hole entropy in different settings. 
For asymptotically Anti-de Sitter eternal ``two-sided'' black holes,     black hole entropy can be explained as the von Neumann entropy of the thermofield double state when restricted to one side of the dual conformal field theory  \cite{Maldacena:2001kr}. In contrast,    ``one-sided''  black holes (with one asymptotic region) correspond  to thermal states in the dual CFT, and the entropy of black holes matches the thermodynamic entropy of the CFT. For instance, the entropy of three-dimensional BTZ black holes   agrees precisely with the Cardy formula for the asymptotic density of states in    2d CFTs \cite{Cardy:1986ie,Strominger:1997eq}. A special case consists of  braneworld holography where the entropy of one-sided black holes is equal to the entanglement entropy across the horizon, since gravity on the brane is   induced by integrating out UV degrees of freedom of the CFT \cite{Emparan:2006ni}. 
We are, however, interested in the bulk quantum gravity theory, and whether the microstates are short range or long range in the bulk spacetime. This is not clear from the dual boundary perspective, as the CFT microstates do not live at a specific location in the bulk.

%The dimensionality of the Hilbert space in quantum gravity in a certain spatial region is bounded by the area of its boundary. This implies a huge redundancy in the Hilbert space. This could be due to a gauge invariance related to diffeomorphism invariance of classical gravity. Another possibility is that the microstates are very non-local, which is indeed suggested by string theory and the recent development on the information paradox. 

%INDUCED GRAVITY
 In induced gravity models black hole entropy is entirely due to short-range  entanglement   across the horizon \cite{Sorkin:2014kta,Frolov:1993ym,Jacobson:1994iw,Frolov:1996aj}. In the vacuum of a relativistic quantum field the von Neumann entropy (a.k.a. entanglement entropy) across a surface with area $A$ is given by \cite{Sorkin:2014kta,Bombelli:1986rw,Srednicki:1993im}
\begin{equation}
S_{vN} (\delta) = \frac{A}{\delta^{d-2}}	 +\, ...
\end{equation}
where $\delta$ is a short distance (UV) regulator. The dots include both  subleading divergent terms and universal finite contributions. In these theories there is no bare gravitational coupling, $1/G_0=0$, which means that gravity   is  induced or emergent. Spacetime is assumed to be present as a background geometry  but its dynamical equation, \emph{i.e.}, the Einstein equation, can be derived by imposing the Clausius relation for local Rindler horizons and invoking the Unruh effect \cite{Jacobson:1995ab,Jacobson:2012yt}.   The value of Newton's constant is determined by the choice of the UV cutoff.   It remains unclear however why, in the framework of quantum field theory, this UV cut off should be fixed at the Planck scale. % It also seems problematic that the semi-classical approximation breaks down at the Planck scale. 

%Quantum field theory, when considered as a continuum theory, has generally too many degrees of freedom, since its entanglement entropy diverges. At the same time quantum field theory has also too little degrees of freedom to account for the entropy of a black hole. Namely, if one restricts to quantum states associated with a quantum field theory inside a given volume, and with a given value of the energy, one generally can not reach the amount of entropy associated with a black hole.  
If we take the induced gravity proposal seriously, then the black hole microstates live at a Planck distance $\ell_P$ away from the horizon. Each Planck cell has an energy given by the Planck mass $M_P = \hbar / \ell_P$. If we add up all the Planck energies of the short range degrees of freedom that account for the black hole entropy, then we arrive at the approximate energy 
%The naive energy of short range degrees of freedoms is
\begin{equation}
	E_{QFT} \sim M_{P} \left (\frac{R_S}{\ell_{P}} \right)^{d-2}\sim   \frac{\hbar}{\ell_P} \left (\frac{R_S}{\ell_{P}} \right)^{d-2}, \label{shortenergy}
\end{equation}
where $R_S$ is the Schwarzschild horizon radius, and the fraction $  ( R_S / \ell_P  )^{d-2}$ is the number of Planck cells on the horizon in $d$ spacetime dimensions. Note that in $d$ dimensions    the Planck length is related to Newton's and Planck's constant by $\ell_{P}^{d-2}= G \hbar$.   We ignore    proportionality factors of order unity, as they depend on the details of the microscopic theory and the choice of UV cutoff, and are not important for the general arguments in this essay.   The answer \eqref{shortenergy} clearly does not agree with the  energy of asymptotically flat Schwarzschild black holes, which is smaller by a factor $\ell_{P} / R_{ S}$ 
\begin{equation}
	E_{BH}  = E_{QFT} \frac{\ell_P}{R_S}  \sim \frac{\hbar}{R_S} \left (\frac{R_S}{\ell_{P}} \right)^{d-2}. \label{longenergy}
\end{equation}
 Note that this energy satisfies the Smarr formula $E_{BH} \sim T_{H} S_{BH} $, where the Hawking temperature is given by $T_H \sim \hbar / R_S$ and the Bekenstein-Hawking entropy is $S_{BH}\sim \left ( R_S/ \ell_P\right)^{d-2}.$ This back-of-the-envelope calculation   makes clear that it is hard to reproduce both the entropy and the energy of black holes simultaneously from a computation in quantum field theory. If one fixes the entropy, then the energy is too high.
If one instead fixes the energy, then quantum field theory has   too few degrees of freedom to account for the entropy of a black hole. Namely, if one restricts to quantum states associated with a quantum field theory inside a given volume, and with a given value of the energy, one generally cannot reach the amount of entropy associated with a black hole. 

%Quantum field theory, when considered as a continuum theory, has generally too many degrees of freedom, since its entanglement entropy diverges. At the same time quantum field theory has also too little degrees of freedom to account for the entropy of a black hole. Namely, if one restricts to quantum states associated with a quantum field theory inside a given volume, and with a given value of the energy, one generally can not reach the amount of entropy associated with a black hole. 

Instead of viewing gravity as being entirely induced, one can also take the point of view that the von Neumann entropy of the quantum fields in spacetime are a subleading correction to the Bekenstein-Hawking area term,  which represents the gravitational contribution. Together they form the generalized entropy, introduced by Bekenstein  to avoid a violation of the second law of thermodynamics \cite{Bekenstein:1972tm,Bekenstein:1973ur}, 
\begin{equation}
	S_{gen} = \frac{A}{4 \hbar G (\delta)}	 + S_{vN} (\delta)  \,.\label{genent}
\end{equation}
%The dots may contain extra Wald entropy terms and counterterms.
There is various evidence that the sum is independent of the UV cutoff~$\delta$, since the divergent terms in the entanglement entropy are canceled by the renormalization of Newton's constant \cite{Susskind:1994sm,Solodukhin:2011gn}. 
In AdS/CFT the first term is to leading order proportional to $N^2$ \cite{Susskind:1998dq}, while the second term scales like $N^0$, where $N$ is the rank of the gauge group $SU(N)$ in the large-$N$ field theory.  The bulk fields do not depend on $N$, and thus can only give an explanation of the second term. This means we still need a microscopic interpretation of the leading $N^2$ term, other than that it corresponds to gravitational entropy. 

In the microscopic theory, which in this case is given by the CFT, it is not clear how to distinguish the two terms in the generalized entropy~\eqref{genent}. % If the QFT fields also live on the Planck scale, then the bulk entanglement entropy also has a $N^2$ term. Deconfinement of boundary fields in AdS/CFT. In AdS/CFT bulk fields have a fall off on the boundary. 
The microscopic interpretation of the full generalized entropy depends on whether one considers an eternal black hole described by a two-sided extended Kruskal spacetime, or a black hole formed by gravitational collapse. In the latter case the black hole formation process may at a microscopic level be described by unitary evolution of an initial pure state. In this case, the black hole entropy measures the number of possible microstates, and hence it represents a coarse grained entropy. An eternal black hole, on the other hand, is believed to be described by the entangled thermofield double state. In this case, the black hole entropy represents a fine grained entropy due to microscopic entanglement.    The coarse grained generalized entropy satisfies the generalized second law of thermodynamics,   while   fine grained entropy is invariant under unitary evolution.

Whether one views the black hole entropy as coarse grained entropy or the entanglement entropy of a thermofield double, in both cases one needs to explain the nature  of the microstates in terms of 
other type of degrees of freedom than the quantum fields in the bulk. These microscopic degrees of freedom should know about the Planck scale, which in the context of AdS/CFT means they should know about $N$. 
The commonly accepted explanation is that for black hole states it no longer suffices to look at “short” single trace operators with a fixed number of matrix valued fields. Instead, one   considers “long” single and multiple trace operators in which the number of matrices grows with $N$. 

In fact, for the counting of microstates it is advantageous to treat the microscopic theory as a version of matrix quantum mechanics instead of a quantum field theory \cite{Banks:1996vh,Dijkgraaf:1997vv}. Each matrix element can be thought of as describing the “ground state” of an open string field: in this interpretation the “long” single trace operators can be interpreted as closed strings whose size and energy grows with $N$. The entropy associated with the black hole microstates is then explained by counting all possible ways in which a state with a given number of matrices with a given size can be written as a product of ``long'' single trace operators. This leads to a version of the Cardy-Hardy-Ramanuyan formula for the entropy in which both the size of the matrix as well as the number of matrix elements needs to be taken into account. The size of the matrix translates into the “central charge” $\mathcal C$ and the number of matrix elements can be denoted by $\mathcal N$. Counting all possible ways in which the $\mathcal N$ elements can be distributed over the $\mathcal C$ degrees of freedom yields the formula for the entropy \cite{Verlinde:2000wg,Majhi:2011ws,Majhi:2012tf,Verlinde:2016toy,Leuven:2018ejp}
\begin{equation}
    S = 4 \pi \sqrt{\mathcal C \mathcal N}.
\end{equation}
We can view the energy as the product of the   number of matrix elements $\mathcal N$ and the excitation energy $\epsilon$ per element $E= \mathcal N \epsilon$.
It turns out that the most entropic states arise when $\mathcal C$ and $\mathcal N$ are of equal size. If we insert $\mathcal N = \mathcal C$ and normalize the size of the matrix as  $\mathcal C (R_S) = A(R_S)/16 \pi G$, then we recover the Bekenstein-Hawking formula $S_{BH} = A(R_S)/4G.$ For black holes whose horizon radius is equal to the AdS radius, $R_S=L$, the dictionary $\mathcal C (L) = A(L)/16 \pi G$ agrees with the Brown-Henneaux formula \cite{Brown:1986nw} for 2d CFTs if $\mathcal C = c/12$, and in general dimensions it agrees with the central charge defined in terms of the normalization of the two-point function of the stress tensor \cite{Myers:2010tj}. A similar dictionary for $\mathcal C(R_S)$ was introduced in \cite{Susskind:1998dq,Leuven:2018ejp}. 

We can   derive the temperature from the Cardy-Hardy-Ramanujan formula as follows
\begin{equation}
	\frac{1}{T} \equiv \left ( \frac{\partial S}{\partial E} \right)_{\mathcal C, \epsilon} = \frac{2\pi }{\epsilon} \sqrt{\frac{\mathcal C}{\mathcal N}}\,  .
\end{equation}
At the horizon we have $\mathcal N = \mathcal C$, hence the horizon temperature is 
\begin{equation}
	T_H = \frac{\epsilon}{2\pi}\, . \label{temp}
\end{equation}
The dictionary for the excitation energy for large AdS black holes $R_S \gg L$    is given by $\epsilon \sim \hbar R_S/L^2$, whereas for small black holes $R \ll L$ it is $\epsilon \sim \hbar/R_S$.  This dictionary is consistent with the expression  for the Hawking temperature for AdS black holes in \cite{Hawking:1982dh}.

 The lowering of the excitation energy in equation \eqref{longenergy} is an example of the long string phenomenon \cite{Leuven:2018ejp,Maldacena:1996ds,Dijkgraaf:1996xw}. This mechanics occurs in large symmetric product CFTs and matrix quantum mechanics, and identifies a twisted sector consisting of ``long strings'' whose typical excitation energy is considerably smaller than that of the untwisted sector.
  In fact, the same long string phenomenon also reduces the number of independent microscopic degrees of freedom. The number of short range degrees of freedom of size $\ell_P$ inside the horizon radius $R_S$ scales with the volume $c_{\ell_P} (R_S)\sim(R_S/\ell_P)^{d-1}$, whereas the number of long range degrees of freedom scales with the area of the horizon 
 \begin{equation}
     \mathcal C = c_{\ell_P} (R_S) \frac{\ell_P}{R_S} \sim \left (\frac{R_S}{\ell_P} \right)^{d-2}.
 \end{equation}
We see that the same factor $\ell_P/ R_S$ lowers both the energy and the number of degrees of freedom. Thus, the long string phenomenon naturally explains why the entropy of black holes is proportional to the area and not the (Euclidean) volume of the horizon. % We refer to   for a more detailed description of the long string mechanism and its relevance for black holes, and for  holography of non-AdS spaces  more generally (like de Sitter holography). % string model of black holes, and also of de Sitter space. 

To conclude, we argued that short range degrees of freedom in quantum field theory cannot account for the entropy and energy of black holes at the same time. In matrix quantum mechanics there are other types of degrees of freedoms, ``long strings", whose typical excitation energy is lower than that of short strings, and the total number of long strings can computed by the Cardy-Hardy-Ramanujan formula reproducing the entropy-area relation for black holes. %\\

\bibliographystyle{JHEP}
\bibliography{longstrings}

\end{document}